\documentclass[sigconf]{acmart}
\AtBeginDocument{%
  }

\setcopyright{acmlicensed}
\copyrightyear{2024}
\acmYear{2024}
\acmDOI{XXXXXXX.XXXXXXX}
\acmConference[WWW'25]{}{April 28--May 02,
  2025}{Sydney, Australia}
\acmBooktitle{WWW '25: International World Wide Web Conference, April 28--May 02, 2025, Sydney, Australia}

\usepackage{array}
\usepackage{tikz}
\def\checkmark{\tikz\fill[scale=0.4](0,.35) -- (.25,0) -- (1,.7) -- (.25,.15) -- cycle;}

\begin{document}


\title{CoinCLIP: A Multimodal Framework for Assessing Viability in Web3 Memecoins}

\author{Hou-Wan Long}
\email{houwanlong@link.cuhk.edu.hk}
\affiliation{%
  \institution{The Chinese University of Hong Kong}
  \city{Shatin}
  \country{Hong Kong}
}

\author{Hongyang Li}
\email{22307140102@m.fudan.edu.cn}
\affiliation{%
  \institution{Fudan University}
  \city{Shanghai}
  \country{China}
}

\author{Wei Cai}
\email{weicaics@uw.edu}
\affiliation{%
  \institution{University of Washington}
  \city{Tacoma}
  \state{Washington}
  \country{USA}
}

\renewcommand{\shortauthors}{Long et al.}

\begin{abstract}
The rapid growth of memecoins within the Web3 ecosystem, driven by platforms like Pump.fun, has made it easier for anyone to create tokens. However, this democratization has also led to an explosion of low-quality or bot-generated projects, often motivated by short-term financial gain. This overwhelming influx of speculative tokens creates a challenge in distinguishing viable memecoins from those that are unlikely to succeed. To address this issue, we introduce CoinVibe, a comprehensive multimodal dataset designed to evaluate the viability of memecoins. CoinVibe integrates textual descriptions, visual content (logos), and community data (user comments, timestamps, and number of likes) to provide a holistic view of a memecoin's potential. In addition, we present CoinCLIP, a novel framework that leverages the Contrastive Language-Image Pre-Training (CLIP) model, augmented with lightweight modules and community data integration, to improve classification accuracy. By combining visual and textual representations with community insights, CoinCLIP provides a robust, data-driven approach to filter out low-quality or bot-driven projects. This research aims to help creators and investors identify high-potential memecoins, while also offering valuable insights into the factors that contribute to their long-term success. The code and dataset are publicly available at \textbf{\hyperlink{https://github.com/hwlongCUHK/CoinCLIP.git}{https://github.com/hwlongCUHK/CoinCLIP.git}}.
\end{abstract}





\maketitle

\section{Introduction}
The rapid expansion of memecoins within the Web3 ecosystem represents a convergence of internet culture and decentralized finance. Unlike traditional cryptocurrencies, memecoins are primarily driven by social media engagement, humor, and viral cultural narratives. Community sentiment plays a crucial role in their success, with memecoins gaining popularity through social trends and celebrity endorsements. Well-known examples such as Dogecoin, Shiba Inu, and Pepe Coin have attracted significant attention, often fueled by viral content and endorsements from public figures. The emergence of platforms like Pump.fun has democratized the creation of memecoins, enabling users to launch tokens without requiring deep technical knowledge, thus fostering a decentralized, community-driven governance model. As of September 2024, Pump.fun has facilitated over 1 million memecoin launches, generating \$100 million in revenue, solidifying its position as a dominant force in the creation of tokens on the Solana blockchain.

However, this low barrier to entry for creating memecoins has also led to the proliferation of low-quality or meaningless tokens. With minimal technical expertise required, both bots and individuals can rapidly deploy memecoins, often with the goal of achieving short-term financial gain. This trend exposes investors to significant risks, as fraudulent or low-value projects are difficult to distinguish from potentially viable ones. Despite the widespread cultural and financial impact of memecoins, existing methodologies for differentiating legitimate projects from those likely to fail or manipulate investors remain underdeveloped.

To address this gap, we introduce \textbf{CoinVibe}, a comprehensive multimodal dataset designed to evaluate the viability of memecoins. CoinVibe integrates textual descriptions, visual content (logos), and community data (user comments, timestamps, and the number of likes), offering a holistic view of memecoin dynamics. This integration enables a more accurate assessment of each memecoin's potential for success by combining content-based and community-driven insights. Additionally, we present \textbf{CoinCLIP}, a novel framework that leverages the Contrastive Language-Image Pre-Training (CLIP) model for multimodal memecoin classification. CoinCLIP enables the simultaneous analysis of textual and visual features, enhanced by community data, to classify memecoins into viable and non-viable categories with greater accuracy.

Our work contributes the following: \textbf{(i)} the introduction of CoinVibe, a multimodal dataset for evaluating memecoin viability, combining textual, visual, and community data; \textbf{(ii)} benchmarking CoinVibe using various unimodal and multimodal methods for memecoin classification; and \textbf{(iii)} the introduction of CoinCLIP, a novel framework that applies lightweight modules on top of a frozen CLIP model to classify memecoins as viable or not viable.

The structure of the paper is as follows: Section 2 introduces Pump.fun and related work. Section 3 describes the CoinVibe dataset. Section 4 presents the CoinCLIP framework. Section 5 discusses the experiment results, including performance metrics and insights. Section 6 concludes the paper.

\begin{figure*}
  \includegraphics[width=\textwidth]{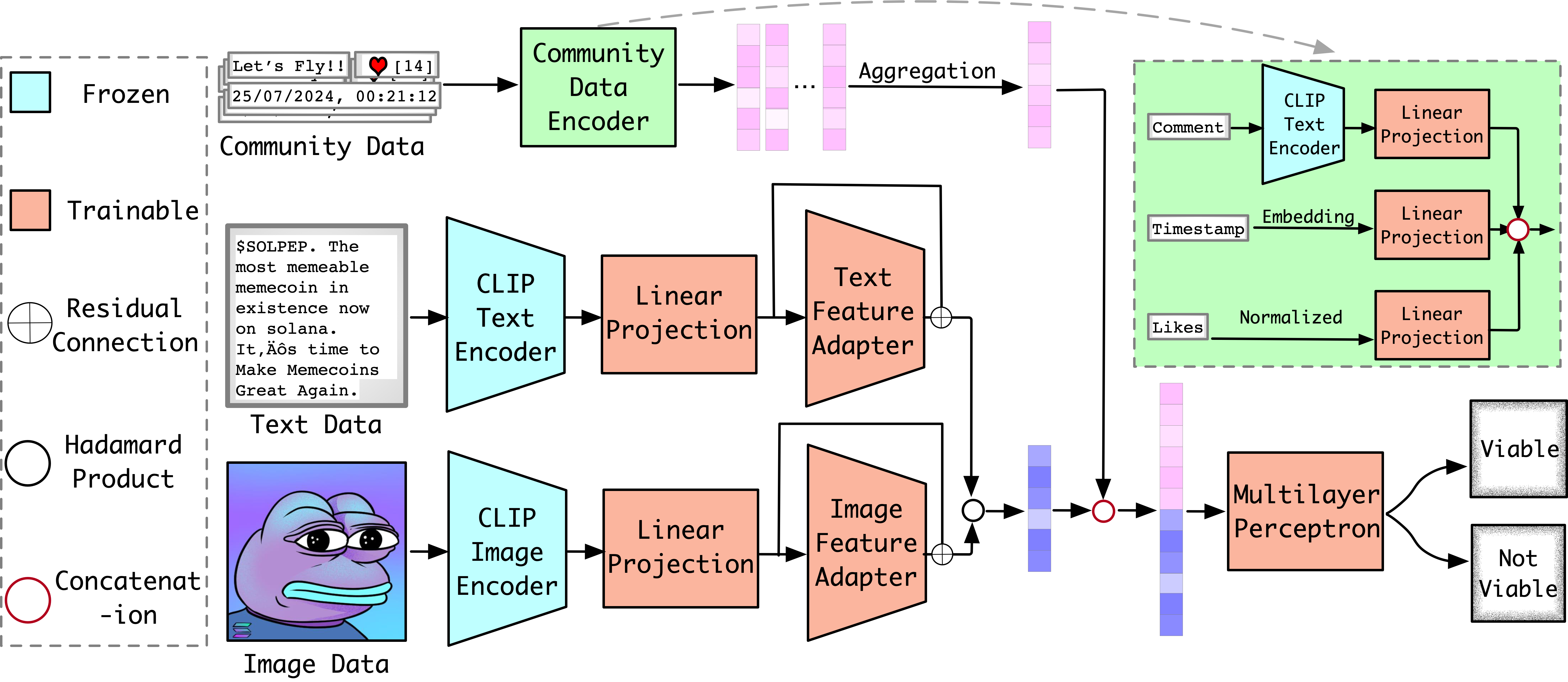}
  \caption{An overview of our proposed framework, CoinCLIP. We use frozen CLIP image and text encoders to create representations for each image-text pair. These representations are passed through linear layers to disentangle the modalities in CLIP’s shared embedding space. We implement Feature Adapters with residual connections for each modality to prevent overfitting. Community data is integrated to enhance the performance.}
  \Description{An overview of our proposed framework, CoinCLIP. We use frozen CLIP image and text encoders to create representations for each image-text pair. These representations are passed through linear layers to disentangle the modalities in CLIP’s shared embedding space. We implement Feature Adapters with residual connections for each modality to prevent overfitting. Community data is integrated to enhance the performance.}
  \label{overview}
\end{figure*}

\section{Pump.fun Memecoins and Related Work}

\text{Pump.fun} is a decentralized platform on the Solana blockchain designed to simplify the creation of memecoins through an easy-to-use interface. The platform allows users to create their own memecoins by defining basic attributes such as the memecoin's name, symbol, and total supply, without requiring any technical expertise. Once created, these tokens attract community engagement through user comments and feedback, which can influence the coin's visibility and growth. If a memecoin's market capitalization reaches \$69,000, it becomes eligible to be listed on Raydium, a decentralized exchange (DEX) on Solana, thereby gaining liquidity and broader exposure within the Solana DeFi ecosystem.

The rise of Web2 platforms has made memes an integral part of internet culture, shaping public opinion, humor, and viral trends. As memes have gained cultural and financial significance, the study of memes through multimodal analysis has become increasingly important. Researchers have explored how meaning is constructed and shared across digital platforms. For instance, Kiela et al. \cite{kiela2020hateful} introduced a multimodal framework to detect hate speech in memes by combining visual and textual embeddings. Similarly, Suryawanshi et al. \cite{suryawanshi2020multimodal} developed the MultiOFF dataset to detect offensive content in memes by analyzing both image and text modalities. Recent works like \cite{shah2024memeclip} by Shah et al. have expanded the scope of multimodal meme analysis to address more complex aspects of expression, including hate, targets of hate, stance, and humor. Additionally, Ling et al. \cite{ling2021dissecting} focused on visual elements that differentiate viral image memes from non-viral ones, and developed a machine learning model to predict meme virality.

However, despite the growing intersection between memes and cryptocurrency in the Web3 ecosystem, research on the viability of memecoins—especially in the context of decentralized platforms like Pump.fun—remains largely underexplored. The proliferation of low-quality or bot-generated memecoins poses a significant challenge in distinguishing legitimate projects from those that are unlikely to succeed. In this study, we build on existing multimodal research to address this challenge, by introducing new methods for evaluating memecoin viability. Our work specifically focuses on the need to distinguish between high-potential memecoins and the large volume of short-lived or fraudulent tokens within the Web3 space.

\section{Dataset Description}

The CoinVibe dataset consists of three primary data components: Textual Descriptions, Visual Content, and Community Data. Textual Descriptions provide narrative information associated with each memecoin, while Visual Content includes the logos of the tokens. Community Data comprises user comments, comment timestamps, and the number of likes each comment received. The dataset covers 6,231 memecoins created between January 2024 and November 2024, offering a comprehensive snapshot of the memecoin ecosystem. Of these, 44.27\% are labeled 'Viable', while the remaining memecoins are labeled 'Non-Viable'.

\subsection{Data Acquisition}

Data was acquired by extracting memecoin metadata from Dune.com using Pump.fun's program address\footnote{6EF8rrecthR5Dkzon8Nwu78hRvfCKubJ14M5uBEwF6P}, which includes memecoin addresses, creation times, names, and creators. Memecoin-specific URLs were generated by appending memecoin addresses to the base URL\footnote{https://pump.fun/}. A web scraping pipeline, built with Selenium and Pandas, was used to automate the extraction of textual descriptions, memecoin images, and user comments. The timestamp and number of likes for each comment were also collected.

\subsection{Data Cleansing}

After data acquisition, a thorough cleaning process was applied to ensure data accuracy. Text data (including descriptions and user comments) were preprocessed by converting to lowercase, removing non-alphabetic characters (e.g., URLs), eliminating stopwords, and applying tokenization and lemmatization. Entries with missing descriptions or comments were excluded. Image files were resized to 224x224 pixels to match the input size for the CLIP Image Encoder, non-RGB images were converted to RGB, and corrupted or missing images were replaced with zero vectors.

\subsection{Data Annotation}

Following data cleansing, the dataset was annotated based on the viability of each memecoin. Viability was determined by whether the memecoin was successfully listed on Raydium, a decentralized exchange (DEX) on the Solana blockchain. Tokens listed on Raydium were labeled as 'Viable', while those not listed were labeled 'Non-Viable'. This classification is based on the premise that being listed on Raydium indicates market validation, strong community engagement, and financial health, making it a reliable indicator of long-term potential. This approach ensures that the dataset is both accurate and relevant for training models to assess the viability of new memecoins.

\section{Methodology}

In this section, we present CoinCLIP, a novel framework for multimodal memecoin classification that leverages the vision-language model CLIP. CLIP generates rich, shared representations of both visual and textual data, making it ideal for memecoin analysis. To tailor CLIP to this task, we introduce lightweight modules that disentangle image and text representations while mitigating overfitting. The overall architecture of CoinCLIP is shown in Figure \ref{overview}. Below, we describe each component in detail.

\textbf{Zero-shot CLIP:} CLIP, pre-trained on 400 million image-text pairs, exhibits strong zero-shot performance. It encodes both visual and textual data into a shared embedding space using its Image and Text Encoders. We freeze the weights of both encoders to retain the valuable pre-trained knowledge, ensuring a strong base for feature extraction.

\textbf{Linear Projection Layers:} To separate image and text representations within CLIP’s shared embedding space, we use individual linear projection layers for each modality. These projections map the image and text features into a consistent format compatible with CLIP’s final hidden state.

\textbf{Feature Adapters:} To prevent overfitting, we incorporate lightweight Feature Adapters for both image and text modalities. These adapters allow the model to fine-tune on new data while preserving CLIP’s pre-trained knowledge. By integrating the outputs of these adapters with the original projections via residual connections, we balance the new features with CLIP’s established knowledge, creating final unimodal representations.

\textbf{Community Data Integration:} We enhance the model by incorporating community data, including user comments, timestamps, and likes. Comments are processed through the CLIP Text Encoder, while timestamps and likes are normalized and transformed into embeddings. These embeddings are linearly projected to align their dimensions and then concatenated into a unified comment representation. To aggregate multiple comments per memecoin, we compute a weighted sum of the comment representations, where each comment's weight is determined by the number of likes it received. The resulting aggregated representation serves as the community data representation.

\textbf{Modality Fusion:} CoinCLIP avoids complex fusion techniques like Cross-Modal Attention by using Hadamard product to combine image and text representations. The resulting multimodal representation is then concatenated with the community data representation, forming a comprehensive feature set for classification.

\textbf{Classification:} A Multilayer Perceptron (MLP) is used for classification. The MLP processes the fused multimodal representation through multiple fully connected layers with ReLU activations, followed by a softmax layer for final prediction. The model is trained using cross-entropy loss and optimized with the Adam optimizer, enabling it to learn complex relationships between the different modalities and classify memecoins accurately.


\section{Experiment}

\subsection{Experiment Setting}

We divide the dataset into train, validation, and test sets with predefined splits of 80\%, 10\%, and 10\%, respectively. Experiments are conducted using both unimodal and multimodal baseline methods, as well as previous frameworks for multimodal meme classification. Each experiment is run with three different random seeds, and the results are reported with the mean and standard deviation ($\pm$) for accuracy, AUC (Macro), and F1-Score (Macro). For all CLIP-based methods, we use the ViT-L/14 model as the image encoder. In the case of CLIP, unimodal feature representations are fused using concatenation.

\subsection{Experiment Result}

\textbf{Unimodal Methods:} For the unimodal methods (see Table \ref{result}), we used ViT-L/14 \cite{dosovitskiy2020image} and CLIP’s image encoder for image-based methods, and BERT \cite{devlin2018bert} and CLIP’s text encoder for text-based methods. The image-based methods generally outperformed the text-based ones, supporting the idea that Transformer-based visual models effectively capture the semantic content of images, including the textual meaning embedded in the visual features \cite{burbi2023mapping}. Specifically, CLIP’s image encoder showed superior performance compared to the standard pre-trained Visual Transformer, despite both models sharing the same architecture, highlighting the effectiveness of contrastive pre-training. Overall, unimodal methods performed worse than all multimodal methods across all tasks, emphasizing the necessity of multimodal processing in memecoin analysis.

\textbf{Multimodal Methods:} We tested several multimodal methods (see Table \ref{result}), including CLIP \cite{radford2021learning}, CLIP-Adapter \cite{gao2024clip}, and our proposed CoinCLIP framework. CoinCLIP outperforms both the baseline CLIP model and the other multimodal methods across all metrics in the viability classification task. While CLIP-Adapter is conceptually similar to CoinCLIP, it uses a single feature adapter, assuming that individual feature adapters for both modalities may introduce redundancy. In contrast, CoinCLIP uses separate feature adapters for both image and text modalities, leading to better performance. This suggests that distinct adapters for each modality are more effective in capturing the unique features of memecoins, where both visual and textual components convey different, yet complementary, meanings.

\begin{table}[ht]
\centering
\begin{tabular}{cccc}
\hline
\hline
\textbf{Method} & \textbf{Acc.} & \textbf{AUC.} & \textbf{F1}\\
\hline
\hline
BERT & $70.17_{\pm 0.63}$  & $75.72_{\pm 0.52}$ & $70.33_{\pm 0.49}$\\
CLIP Text-Only & $72.16_{\pm 0.36}$ & $78.82_{\pm 0.66}$ & $71.60_{\pm 0.21}$\\
ViT-L/14 & $76.92_{\pm 1.23}$ & $84.28_{\pm 0.49}$ & $73.27_{\pm 1.92}$\\
CLIP Image-Only & $79.12_{\pm 0.24}$ & $88.32_{\pm 2.17}$ & $78.72_{\pm 0.32}$\\
\hline
CLIP & $81.36_{\pm 0.81}$ & $87.27_{\pm 0.67}$ & $80.30_{\pm 0.95}$\\
CLIP-Adapter & $82.21_{\pm 0.73}$ & $87.53_{\pm 0.61}$ & $80.89_{\pm 0.87}$\\
\hline
CoinCLIP & $\textbf{84.72}_{\pm 0.45}$ & $\textbf{92.07}_{\pm 0.34}$ & $\textbf{83.74}_{\pm 0.43}$\\
\hline
\hline
\end{tabular}
\caption{Classification performance of CoinCLIP and baseline methods on the CoinVibe dataset, reported as Mean $\pm$ Standard Deviation across three evaluation metrics: Accuracy, AUROC (Macro), and F1-Score (Macro). The best performance is highlighted in bold.}
\label{result}
\end{table}

\subsection{Ablation Study}

We conducted a systematic ablation study to evaluate the contribution of each component of CoinCLIP to its overall performance. The results of our ablation experiments are presented in Table \ref{ablation}. We began with the CLIP ViT-L/14 model and progressively integrated each external module of CoinCLIP. The introduction of projection layers significantly improved performance, demonstrating the importance of adapting CLIP’s embedding spaces for the downstream task by disentangling image and text representations. Initially, adding Feature Adapters caused a slight dip in F1-Score, but over time, these adapters allowed the model to refine image and text representations by incorporating learnable parameters. Integrating community data representation further enhanced performance, indicating that community engagement metrics—such as user comments, timestamps, and likes—contribute valuable context for memecoin classification. This approach further improved generalization, enabling the model to better capture subtle distinctions between viable and non-viable memecoins.

\begin{table}[t]
\centering
\begin{tabular}{ccccccc}
\hline
\hline
\textbf{CLIP} & \textbf{PL} & \textbf{FA} & \textbf{CDI} & \textbf{Acc.} & \textbf{AUC.} & \textbf{F1} \\
\hline
\hline
\checkmark &  &  &  & $71.23_{\pm 0.53}$ & $79.34_{\pm 0.98}$ & $71.23_{\pm 1.13}$ \\
\checkmark & \checkmark &  &  & $73.52_{\pm 0.13}$ & $80.45_{\pm 0.29}$ & $74.89_{\pm 0.56}$ \\
\checkmark & \checkmark & \checkmark &  & $75.34_{\pm 0.29}$ & $83.39_{\pm 0.11}$ & $73.46_{\pm 0.44}$ \\
\hline
\checkmark & \checkmark & \checkmark & \checkmark & $\textbf{76.44}_{\pm 0.19}$ & $\textbf{84.07}_{\pm 0.19}$ & $\textbf{74.98}_{\pm 0.12}$ \\
\hline
\hline
\end{tabular}
\caption{Ablation experiments conducted on CoinCLIP using the hate detection task from the CoinVibe dataset. Results are reported as Mean $\pm$ Standard Deviation. PL, FA, and CDI refer to Projection Layers, Feature Adapters, and Community Data Integration, respectively. The final row represents the complete CoinCLIP framework. The best-performing results are highlighted in bold.}
\label{ablation}
\end{table}

\section{Conclusion}
In this paper, we introduce CoinVibe, a comprehensive multimodal dataset designed to evaluate the viability of memecoins by integrating textual descriptions, visual content, and community-driven insights. We also present CoinCLIP, a novel framework that leverages the power of CLIP’s image-text representations, enhanced by community data, to classify memecoins into viable and non-viable categories. Through extensive experiments and comparison with baseline models, we demonstrate that CoinCLIP outperforms existing approaches, achieving superior accuracy, AUROC, and F1-Score. Our ablation study further highlights the significance of each component in CoinCLIP, with features such as projection layers, feature adapters, community data integration, and semantic-aware initialization all contributing to improved performance.

The results of this work offer valuable insights into the multimodal nature of memecoin classification and provide a strong foundation for future research in this area. As memecoins continue to play an increasingly important role in the Web3 ecosystem, tools like CoinCLIP can help both investors and developers navigate the challenges posed by this dynamic and rapidly evolving space. Moving forward, we envision the integration of additional features, such as  network-based indicators, to further enhance the predictive capabilities of memecoin classification models.




\bibliographystyle{ACM-Reference-Format}
\bibliography{sample-base}


\end{document}